\documentclass[10pt,journal,compsoc]{IEEEtran}
\ifCLASSOPTIONcompsoc
  \usepackage[nocompress]{cite}
\else
  \usepackage{cite}
\fi
\ifCLASSINFOpdf
\else
\fi
\usepackage{amsmath,amsfonts}
\usepackage{array}
\usepackage[caption=false,font=normalsize,labelfont=sf,textfont=sf]{subfig}
\usepackage{textcomp}
\usepackage{stfloats}
\usepackage{url}
\usepackage{verbatim}
\usepackage{graphicx}
\usepackage{cite}
\usepackage{amsmath}

\usepackage[linesnumbered,ruled,vlined]{algorithm2e}
\usepackage{multirow}
\usepackage{pifont}
\usepackage{booktabs}
\usepackage{xcolor}
\usepackage{soul}
\usepackage{amssymb}

\usepackage[bookmarks=true,breaklinks=true,colorlinks,linkcolor=blue,citecolor=blue,urlcolor=black]{hyperref}

\newcommand{\squishlist}{
   \begin{list}{$\bullet$}
    { \setlength{\itemsep}{0pt}      \setlength{\parsep}{0pt}
      \setlength{\topsep}{3pt}       \setlength{\partopsep}{0pt}
      \setlength{\listparindent}{-2pt}
      \setlength{\itemindent}{-5pt}
      \setlength{\leftmargin}{1em} \setlength{\labelwidth}{0em}
      \setlength{\labelsep}{0.5em} } }

\newcommand{\squishend}{
    \end{list}  }

\setuldepth{cat}

\newcommand{\note}[1]{{\color{magenta}$\square$}}

\begin{document}
%
\title{SiHGNN: Leveraging Properties of \underline{S}emant\underline{i}c Graphs for Efficient \underline{HGNN} Acceleration}
%
%
%
%

\author{Runzhen~Xue,
        Mingyu~Yan,~\IEEEmembership{Member,~IEEE},
        Dengke~Han,
        Ziheng~Xiao,\\
        Zhimin~Tang,
        Xiaochun~Ye,
        and~Dongrui~Fan,~\IEEEmembership{Senior~Member,~IEEE}
\IEEEcompsocitemizethanks{
\IEEEcompsocthanksitem 
Runzhen Xue, Mingyu Yan, Dengke Han, Zhimin Tang, Xiaochun Ye, and Dongrui Fan are with the State Key Lab of Processors, Institute of Computing Technology, Chinese Academy of Sciences, Beijing 100045,
China, and also with the University of Chinese Academy of Sciences, Beijing 101408, China. 
Zhimin Tang is also with the Faculty of Computility Microelectronics, Shenzhen University of Advanced Technology, Shenzhen 518107, China.
E-mail: \{xuerunzhen21s, yanmingyu, handengke21s, tang, yexiaochun, fandr\}@ict.ac.cn. 

\IEEEcompsocthanksitem 
Ziheng Xiao is with the State Key Lab of Processors, Institute of Computing Technology, Chinese Academy of Sciences, Beijing 100045,
China
E-mail: xiaoziheng99@gmail.com.

\IEEEcompsocthanksitem 
Mingyu Yan is the corresponding author of this paper.
}

\thanks{Manuscript received April 19, 2005; revised August 26, 2015.}}

%
%

\markboth{IEEE Transactions on Parallel and Distributed Systems}%
{Xue \MakeLowercase{\textit{et al.}}: SiHGNN}
%



\IEEEtitleabstractindextext{%

\begin{abstract}

Heterogeneous Graph Neural Networks (HGNNs) have expanded graph representation learning to heterogeneous graph fields.
Recent studies have demonstrated their superior performance across various applications, including medical analysis and recommendation systems, often surpassing existing methods. 
However, GPUs often experience inefficiencies when executing HGNNs due to their unique and complex execution patterns. Compared to traditional Graph Neural Networks, these patterns further exacerbate irregularities in memory access. To tackle these challenges, recent studies have focused on developing domain-specific accelerators for HGNNs. Nonetheless, most of these efforts have concentrated on optimizing the datapath or scheduling data accesses, while largely overlooking the potential benefits that could be gained from leveraging the inherent properties of the semantic graph, such as its topology, layout, and generation.

In this work, we focus on leveraging the properties of semantic graphs to enhance HGNN performance. 
First, we analyze the Semantic Graph Build (SGB) stage and identify significant opportunities for data reuse during semantic graph generation. 
Next, we uncover the phenomenon of buffer thrashing during the Graph Feature Processing (GFP) stage, revealing potential optimization opportunities in semantic graph layout.
Furthermore, we propose a lightweight hardware accelerator frontend for HGNNs, called SiHGNN. This accelerator frontend incorporates a tree-based Semantic Graph Builder for efficient semantic graph generation and features a novel Graph Restructurer for optimizing semantic graph layouts. 
Experimental results show that SiHGNN enables the state-of-the-art HGNN accelerator to achieve an average performance improvement of 2.95$\times$. 


\end{abstract}


}

\maketitle

\section{Introduction}
\IEEEPARstart{G}{raph} Neural Networks (GNNs) have drawn tremendous attention in the past few years due to their unique ability to extract latent information from graph data~\cite{gnn_algorithm_survey,gnn_distributed_training_survey}. However, the earlier success of GNNs is mostly concentrated on homogeneous graphs (HomoGs), i.e., graphs with only one type of vertices and edges. While, many real-world data in complex systems are naturally represented as heterogeneous graphs (HetGs), which consist of multiple types of entities and relations that are embodied by various types of vertices and edges, respectively. Compared to HomoGs, HetGs possess not only structural information but also rich semantic information~\cite{HG_survey}. 
Due to the powerful representational ability of HetG, it has been widely adopted to model relational data across various critical fields, including circuit~\cite{GCN-RL, ironman_pro}, knowledge graphs~\cite{oh2018knowledge, socher2013reasoning, zhang2019iteratively, chen2017task}, social networks~\cite{yasunaga2019scisummnet,zhou2015cross,tajeuna2018modeling,zheng2020clustering}, and many others.

Heterogeneous Graph Neural Networks (HGNNs) have been developed to effectively capture both the structural and semantic information inherent in HetGs.
They have reportedly been achieving state-of-the-art performance in various critical applications including recommendation systems~\cite{li2022disentangled}, knowledge inference~\cite{CompGCN}, information retrieval~\cite{mao2020item}, medical analysis~\cite{luo2021imas}, etc.
Within HGNNs, transductive learning and inductive learning represent two distinct approaches to generalizing from data~\cite{graphsage}. Transductive learning involves training a model on a specific graph and making predictions for that same graph. In contrast, inductive learning trains a model on one or more graphs and applies it to make predictions on entirely new, unseen graphs.

The inference process of most inductive HGNN models typically consists of two stages: the Semantic Graph Building (SGB) stage and the Graph Feature Processing (GFP) stage. During the SGB stage, the original HetG is partitioned into multiple semantic graphs based on predefined rules, laying the groundwork for the subsequent processing in the GFP stage.
The GFP stage is further divided into three sub-stages~\cite{HiHGNN, understand_HGNN, GNN_characterization_survey,MetaNMP}: the \textit{Feature Projection} (FP) sub-stage, which transforms the feature vector of each vertex in each semantic graph using a multi-layer perceptron; the \textit{Neighbor Aggregation} (NA) sub-stage, which aggregates features from neighbors for each vertex in each semantic graph; and the \textit{Semantic Fusion} (SF) sub-stage, which fuses the semantic information from all semantic graphs by combining the results of the NA sub-stage across different semantic graphs for each vertex.

Due to the unique workflow outlined above, current hardware solutions, such as GPUs and GNN accelerators, face challenges in efficiently executing HGNNs~\cite{HiHGNN, understand_HGNN,GNN_characterization_survey,HGNN_training_profiling,GDR-HGNN,MetaNMP,ADE-HGNN}.
GPUs, for instance, struggle with efficiently handling irregular memory accesses stemming from the graph-topology-dependent program behavior in the GFP stage~\cite{HiHGNN, understand_HGNN,GNN_characterization_survey,HGNN_training_profiling,GDR-HGNN,MetaNMP,ADE-HGNN}. On the other hand, GNN accelerators~\cite{HyGCN,GraphACT,AWB_GCN,igcn,FlowGNN, multigcn_inter_node_communication} tailored their hardware datapath and corresponding control to GNNs, such as HyGCN~\cite{HyGCN}, lack the HGNN-oriented control and execution units to process the unique workflow of HGNNs~\cite{HiHGNN, understand_HGNN}.

In response to these challenges, recent work~\cite{HiHGNN, MetaNMP, ADE-HGNN, GDR-HGNN} has focused on the HGNN acceleration.
For instance, to achieve stage fusion during the inference phase and harness the potential parallelism between the processing of semantic graphs, HiHGNN~\cite{HiHGNN} employs a multi-lane architecture. It also strategically schedules the execution order of semantic graphs based on their similarity, thereby exploiting data reusability during execution.
Despite significant advancements, most efforts have focused on datapath optimization or data access scheduling, overlooking potential opportunities related to the properties of the semantic graph itself, such as its topology, layout, and generation.

In this work, we begin our exploration by scrutinizing the characteristics in the execution of HGNNs, uncovering opportunities to leverage the properties of semantic graphs for accelerating HGNNs. We first investigate the SGB stage, identifying significant opportunities for data reuse during semantic graph generation. Next, we expose the issue of buffer thrashing during the GFP stage, highlighting potential optimization opportunities in the layout of semantic graphs.
To further capitalize on these opportunities, we propose a lightweight accelerator frontend called SiHGNN, intended for integration into existing HGNN accelerators to enhance performance. 
This accelerator frontend comprises a tree-based Semantic Graph Builder and incorporates a novel Graph Restructurer.
For the former, SiHGNN introduces a novel Callback Tier Tree (CTT) technique that optimizes semantic graph generation by reusing intermediate results, thereby eliminating redundant computation and memory access in the SGB stage.
For the latter, SiHGNN proposes a graph restructuring algorithm that separates the original semantic graph into a set of edges that do not share common vertices and restructures them into a series of subgraphs, each characterized by a robust community structure. This approach improves data locality and reduces memory overhead, leading to enhanced performance and efficiency.

To summarize, we list our contributions as follows:
\squishlist

\item We examine the execution properties of semantic graph generation in the SGB stage, revealing opportunities for data reuse during the generation of semantic graphs.

\item We identify the occurrence of buffer thrashing during the GFP stage, revealing potential opportunities to optimize the layouts of semantic graphs by leveraging their inherent properties.

\item We propose SiHGNN, a lightweight hardware accelerator frontend designed to enhance the execution of HGNNs. SiHGNN features a tree-based Semantic Graph Builder for efficient semantic graph generation and a Graph Restructurer for optimizing semantic graph layout.

\item Experimental results demonstrate that SiHGNN enables the state-of-the-art HGNN accelerator to achieve an average performance improvement of 2.95$\times$ with only a 2.80\% increase in total area and a 0.68\% increase in total power.


\squishend

\section{Background}\label{sec:background}

In this section, we introduce the pertinent concepts for HGNNs.
Table~\ref{tb:notation} gives all notations used in this work.

\begin{table}[!t]
\centering
\caption{Notations and Corresponding Explanations.}
\label{tb:notation}
\resizebox{0.45\textwidth}{!}{
\tabcolsep=0.5pt
\begin{tabular}{cc|cc}
\toprule
\textbf{Notation}                & \textbf{Explanation}                           & \textbf{Notation}                  & \textbf{Explanation}       \\ \midrule
$G$                     & heterogeneous graph                   & $V$                       & vertex set \\
$V_{src}(V_{dst})$      & source (dest) vertex set & $E$                     & edge set    \\ 
$u,\,v$                   & vertex & e ($e_{u,v}$)           & edge (from $u$ to $v$)   \\
$N(v)$ & neighbor vertex set of $v$  & $\mathcal{T}^v$           & vertex type set \\
$\mathcal{T}^e$         & edge type set         & $G^{\mathcal{P}}$         & semantic graph \\
$G_{s}^{\mathcal{P}}$     & subgraphs restructured by $G^{\mathcal{P}}$  & $\mathcal{R}$    & relation    \\
\bottomrule
\end{tabular}}
\end{table}

\subsection{Heterogeneous Graph}
A graph can be defined as $G=(V, E, \mathcal{T}^v, \mathcal{T}^e)$, where $V$ is the vertex set, $E$ is the edge set, $\mathcal{T}^v$ is the vertex type set and $\mathcal{T}^e$ is the edge type set. A graph is HetG when $|\mathcal{T}^v|+|\mathcal{T}^e|>2$. 
In HetG, each edge type is termed as a relation $\mathcal{R} \in \mathcal{T}^e$, while an edge $e_{u,v}\in E$ starts from the source vertex $u$ and ends at the target vertex $v$. 
For example, the relation A $\rightarrow$ M in the IMDB dataset means that actor A acts in a movie M. 

A \textbf{semantic graph} is generated by different metapaths. A metapath $\mathcal{P}$ defines a composite relation of several edge types, represented as a semantic path in the form of $\mathcal{P} = c_1\rightarrow c_2\rightarrow \ldots\rightarrow c_{l}$ ($\mathcal{P}= c_1c_2\ldots c_{l}$ for short). 
Due to the metapath refers to a sequence of vertex types and edge types that captures specific semantic relationships between vertices, it provides a higher-level abstraction of the graph structure and enables the identification of meaningful paths and patterns within the graph, such as author$\rightarrow$paper$\rightarrow$author (abbreviated as APA), which represents a coauthor relationship. In this way, a relation is the basic semantic graph considered as a one-hop metapath. By utilizing metapaths, we can gain insights into the complex relationships and dependencies present in HetGs.

\subsection{Heterogeneous Graph Neural Networks} 
\begin{figure}[!h] 
	\centering
	\includegraphics[width=0.49\textwidth]{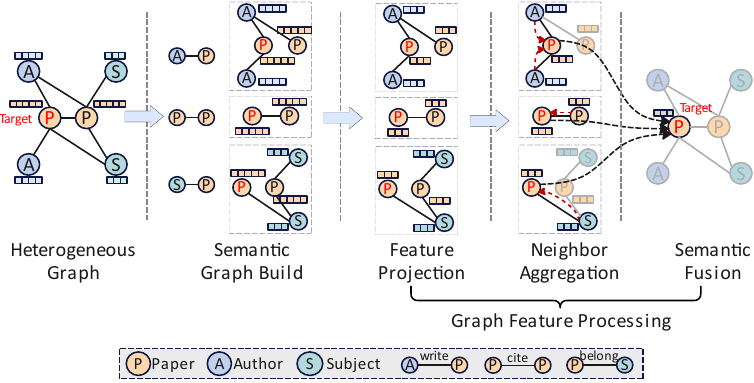}
	\caption{An example of HetGs and execution process of HGNN models.}
	\label{fig:HGNN_work_flow}
\end{figure}

HGNN follows a neighborhood aggregation scheme and a semantic fusion scheme, where the final representation of each vertex is computed by recursively aggregating the feature vectors of its neighbor vertices in each semantic graph and fusing the aggregated results across all semantic graphs, as shown in Fig.~\ref{fig:HGNN_work_flow}.
For example, HAN~\cite{HAN} aggregates structural information using the neighbor attention in each semantic graph and then fuses outputs from different semantic graphs using the semantic attention for each vertex.

To capture both the structural information and semantic information in HetGs, the most prevalent HGNN models usually contain two major execution stages as shown in Fig.~\ref{fig:HGNN_work_flow}.
\ding{172} \textbf{Semantic Graph Build:} The SGB stage builds semantic graphs for the following stages by partitioning the original HetG into a set of semantic graphs based on predefined metapaths.
\ding{173} \textbf{Graph Feature Processing:} This stage processes and integrates features within each semantic graph and can be further divided into three sub-stages~\cite{sehgnn,understand_HGNN,HiHGNN}:
\textit{Feature Projection:} In the FP sub-stage, the feature vector of each vertex is transformed into a new one using a multi-layer perceptron within each semantic graph.
\textit{Neighbor Aggregation:} The NA sub-stage utilizes an attention mechanism~\cite{GAT} to perform a weighted sum aggregation of features from neighbors within each semantic graph.
\textit{Semantic Fusion:} The SF sub-stage fuses the semantic information obtained from all semantic graphs with an attention mechanism~\cite{GAT}, aiming to combine the results of the NA sub-stage across different semantic graphs for each vertex.

\subsection{Differences between GNNs and HGNNs}

The major differences between GNNs and HGNNs on execution are as follows~\cite{understand_HGNN,HGNN_training_profiling,HiHGNN,MetaNMP}.

$\bullet$ \textbf{Direct vs. Indirect Graph Processing:}
GNNs operate directly on given HomoGs, leveraging the graph structure to propagate information. In contrast, HGNNs do not execute directly on complete HetGs; instead, they transform them into semantic graphs before execution.

$\bullet$ \textbf{Joint vs. Separate Feature Projection}: In HomoGs, vertices share the same vector space and dimension, allowing for joint feature projection. In HetGs, vertices of different types have distinct vector spaces and dimensions, necessitating separate feature projection parameters. 

$\bullet$ \textbf{Aggregation vs. Aggregation+Fusion}: GNNs aggregate features once for neighbor aggregation on a single type of relation. In contrast, HGNNs aggregate features from neighbors in each semantic graph and then fuse intermediate results of each semantic graph for each vertex.

\section{Motivation}\label{sec:motivation}
This section highlights the inefficiencies in the SGB and GFP stages, which serve as the motivation behind our design.

\subsection{Low Efficiency in Semantic Graph Generation}
This subsection analyzes the execution of the SGB stage, revealing the low efficiency in semantic graph generation.

Unlike traditional GNNs~\cite{GCN_profiling,GNN_characterization_survey,GNN_architectural_implications_GPU_TPU,distributed_gnn_training_gpu}, HGNNs do not perform inference directly on the complete HetG. Instead, they first transform the HetG into semantic graphs during the SGB stage and then independently process these semantic graphs until the SF sub-stage. Additionally, previous work~\cite{sehgnn} has shown that leveraging longer metapaths can capture more semantic information, thereby enhancing prediction accuracy in HGNN models.

\begin{figure}[!h] 
	\centering
	\includegraphics[width=0.48\textwidth]{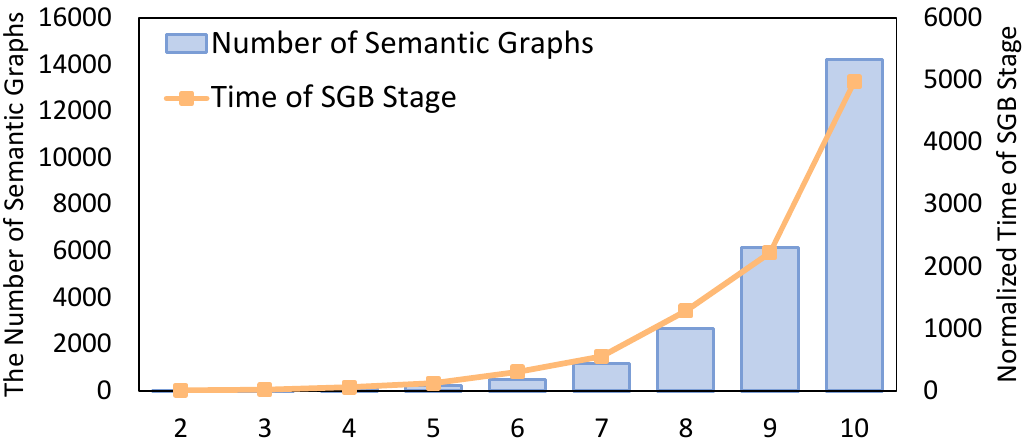}
	\caption{The number of semantic graphs and the normalized time of the SGB stage across various lengths of the metapaths.}
	\label{fig:motivation_SGB}
        \vspace{-10pt}
\end{figure}

The conventional method of generating semantic graphs involves incrementally concatenating relations until the target metapath is fully formed. For example, to construct the metapath author$\rightarrow$paper$\rightarrow$author (APA), the relations author$\rightarrow$paper (AP) and paper$\rightarrow$author (PA) are combined. This approach is effective and straightforward for relatively short metapaths. However, as metapaths lengthen, the number of involved relations increases, leading to different long metapaths potentially sharing the same generation steps. 
Fig.~\ref{fig:motivation_SGB} illustrates how the total number of semantic graphs and the time required for the SGB stage increase with longer metapaths in the ACM dataset.
For instance, to create a four-hop metapath like APSPA, the traditional method concatenates four shorter metapaths: AP-PS-SP-PA. However, generating another metapath, APSPP, requires concatenating AP-PS-SP-PP, where AP-PS-SP is used twice, leading to redundant computation and memory access. 
This issue is further exacerbated as the metapath lengthens. 



\subsection{Low Cache Hit Rate and Buffer Thrashing}
This subsection quantitatively analyzes the low cache hit rate on the GPU platform and the buffer thrashing issue in the HGNN accelerator, which highlights the need for the SiHGNN design. These challenges significantly impede HGNN acceleration efficiency. To investigate further, we conduct a quantitative analysis using the NVIDIA T4 GPU and the state-of-the-art HGNN accelerator, HiHGNN~\cite{HiHGNN}

\begin{figure}[!t] 
	\centering
	\includegraphics[width=0.48\textwidth]{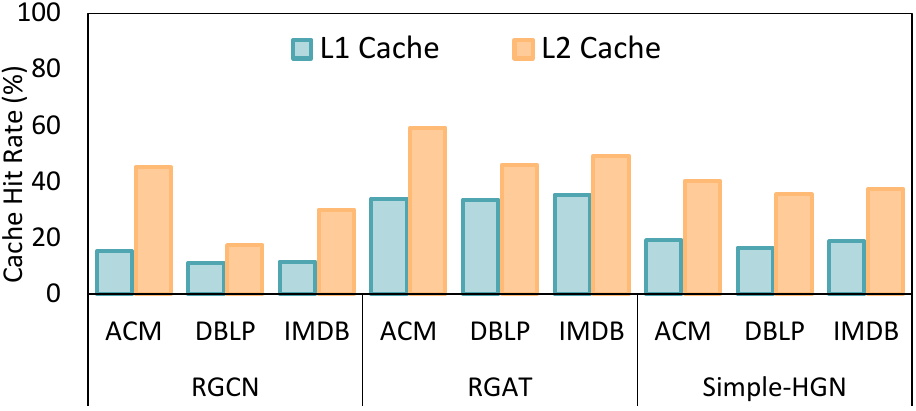}
	\caption{Analysis on T4 GPU with various models: The L1 and L2 cache hit rate during NA sub-stage.}
	\label{fig:motivation1}
\end{figure}

We conduct a quantitative experiment for the NA sub-stage of the RGCN~\cite{RGCN} model using a state-of-the-art framework, DGL~\cite{DGL}, running on an NVIDIA T4 GPU. 
As shown in Fig.~\ref{fig:motivation1}, the experimental results reveal suboptimal L1 and L2 cache hit rates during processing. For instance, with the IMDB and DBLP datasets, the L2 cache hit rate reaches only 30.1\% and 17.5\% in the NA sub-stage, respectively. This indicates that a significant number of vertex features undergo frequent replacements, particularly since the NA sub-stage dominates HGNNs, accounting for up to 74\% of the total GFP stage time~\cite{understand_HGNN, HGNN_training_profiling}. This issue arises because the NA sub-stage aggregates features from neighboring vertices based on the irregular topology of semantic graphs, resulting in irregular memory accesses that degrade overall performance. These findings underscore the need for more efficient memory access patterns in HGNNs to address these challenges.

\begin{figure}[!h] 
	\centering
	\includegraphics[width=0.48\textwidth]{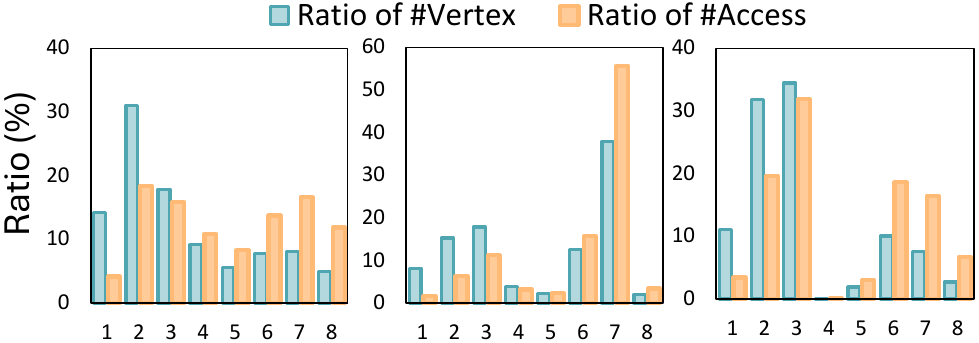}
	\caption{Analysis of HiHGNN with the RGCN model: Frequency of vertex feature replacements during the NA sub-stage. The datasets, from left to right, are ACM, DBLP, and IMDB.}
	\label{fig:motivation2}
        \vspace{-10pt}
\end{figure}

Fig.~\ref{fig:motivation2} gives statistics on the replacement times of vertex features from the buffer during the NA sub-stage on HiHGNN.
The numbers on the horizontal axis represent the replacement times of vertices' features, while the ``Ratio of \#Vertex'' represents the ratio of the number of vertices with specific replacement times to the total number of vertices. Similarly, ``Ratio of \#Access'' denotes the ratio of the number of DRAM accesses conducted by vertices with specific replacement times to the total number of DRAM accesses. The results indicate that a considerable number of vertex features undergo frequent replacements, contributing to the buffer thrashing issue and resulting in a substantial number of redundant DRAM accesses. This excessive data movement significantly hinders overall performance.
It's worth noting the varying degrees of buffer thrashing across the three datasets, attributed to their different graph sizes. The DBLP dataset exhibits the most pronounced occurrence, primarily due to its significantly larger number of vertices compared to the other datasets.

\section{Architecture Design} 
This section presents the design of SiHGNN. We first provide an overview of SiHGNN and then describe the Semantic Graph Builder and Graph Restructurer in detail.

\subsection{Overview of SiHGNN}

\begin{figure}[!hptb] 
	\centering
	\includegraphics[width=0.48\textwidth]{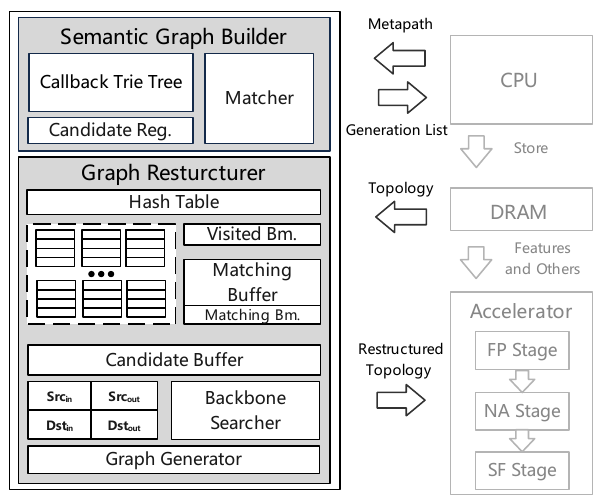}
	\caption{Design and workflow overview of SiHGNN.}
	\label{fig:overview}
                \vspace{-10pt}
\end{figure}

Fig.~\ref{fig:overview} illustrates an overview of SiHGNN, which consists of two main components: the Semantic Graph Builder and the Graph Restructurer. The Semantic Graph Builder is responsible for optimizing the generation of the candidate semantic graph in the SGB stage, eliminating redundant computations and memory accesses. The Graph Restructurer restructures graph topology before the GFP stage, strengthens the community structure of semantic graphs, and alleviates buffer thrashing during the NA sub-stage.

During HGNN inference, the CPU first sends the metapath for the semantic graph that needs to be constructed to the Semantic Graph Builder. Leveraging the CTT, the Semantic Graph Builder then provides the CPU with the optimal generation list, outlining the combination of short metapaths required for its creation. This approach maximizes the reuse of previously generated semantic graphs, ensuring efficient construction.
Once the semantic graph is built, the DRAM transfers the topology to the Graph Restructurer.
The Graph Restructurer decouples the original semantic graph using the decoupler and restructures it with the recoupler. By establishing a dataflow between SiHGNN and other accelerators, efficient HGNN execution is achieved through the pipelined datapath. Both SiHGNN and the combined accelerators operate concurrently and share the memory controller to manage data interactions between on-chip buffers and high-bandwidth memory (HBM).

\subsection{Semantic Graph Builder}
In this section, we provide detailed information about the Semantic Graph Builder, including the introduction of the CTT algorithm and its hardware implementation.

\subsubsection{Callback Trie Tree}
As mentioned in Section~\ref{sec:motivation}, there are redundant computations and memory accesses in the SGB stage, especially as the metapath lengthens. To address this issue, we propose the CTT, which guides the generation of semantic graphs and eliminates redundancies by enabling metapaths reuse.

\begin{figure*}[!hptb] 
	\centering
	\includegraphics[width=0.95\textwidth]{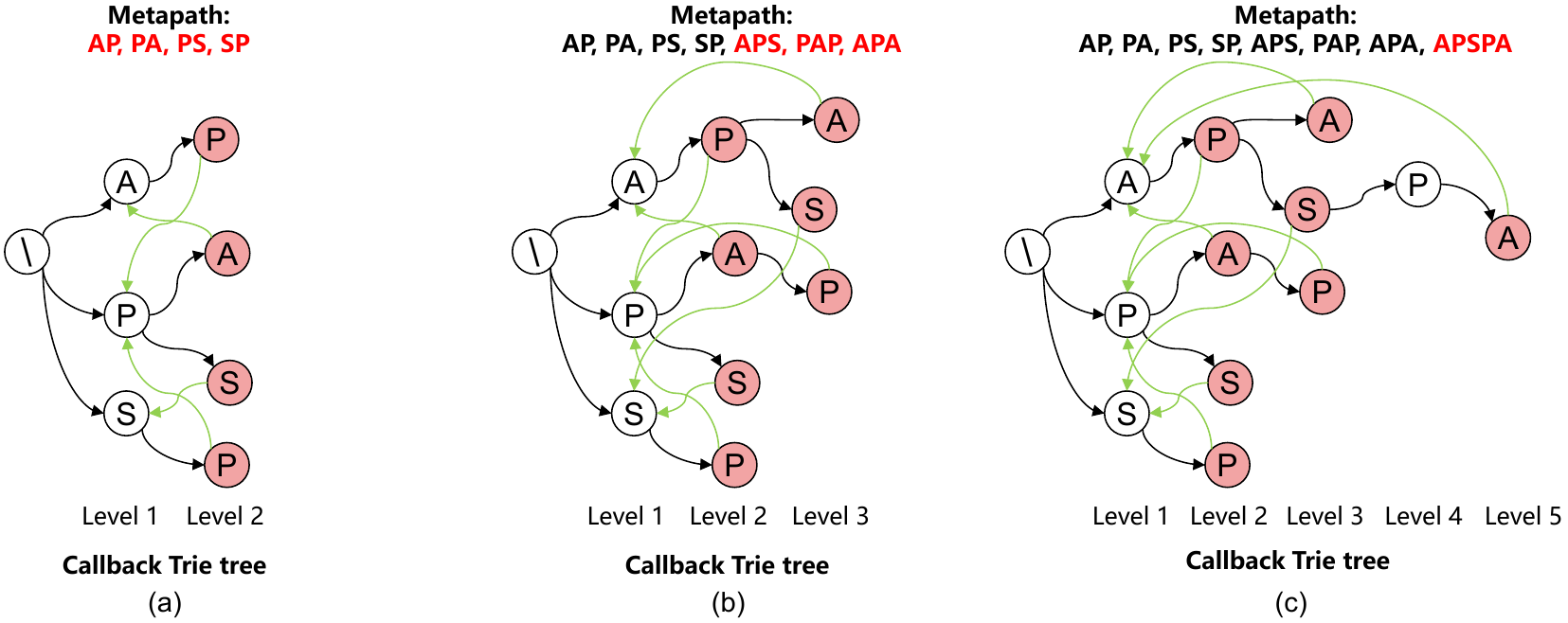}
                \vspace{-10pt}
	\caption{An example of building a Callback Trie Tree. (a) Initialize the Callback Trie Tree with metapaths: AP, PA, PS, SP; (b) Generate the new metapaths: APS, PAP, APA; (c) Generate the new metapath: APSPA.}
	\label{fig:CTT}
\end{figure*}

Fig.~\ref{fig:CTT} illustrates how the CTT guides the generation of semantic graphs. Initially, we start with one-hop metapaths (relations) and build the trie tree shown in Figure~\ref{fig:CTT}(a). Next, we add a callback edge (green edges) for each leaf node (red nodes), pointing to the same node in the level 1 node as itself, resulting in a callback trie tree.
Next, for semantic graphs obtained from two-hop metapaths, such as APA, the CTT first traverses from node A at level 1 and searches node P at level 2, where it finds AP. At this point, CTT reaches a leaf node, and the first sub-metapath for generating APA is identified as AP. Using the callback edge, CTT then traverses to node P at level 1. Following this, node P finally finds node A at level 2, as shown in Figure~\ref{fig:CTT}(b). Thus, the metapath APA can be generated from the sub-metapaths AP and PA using the CTT, and the same applies to APS and PAP.
When generating semantic graphs from multi-hop metapaths, CTT leverages existing short metapaths to construct longer ones, such as APSPA. Through CTT, APSPA can be generated using the sub-metapaths APS, SP, and PA, as shown in Figure~\ref{fig:CTT}(c), rather than starting from one-hop relations.

\subsubsection{Hardware Implementation}
\begin{figure}[!hptb] 
	\centering
	\includegraphics[width=0.48\textwidth]{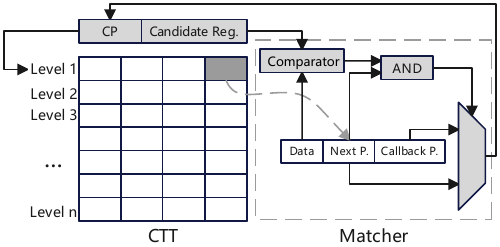}
	\caption{Micro-architecture of Semantic Graph Builder.}
	\label{fig:builder}
\end{figure}

Fig.~\ref{fig:builder} depicts the mapping of Semantic Graph Builder to hardware implementation.
The Semantic Graph Builder consists of the register files, the CTT buffer, and the Matcher.

In the SGB stage, metapaths obtained from one-hop metapaths are first stored in the CTT buffer, forming a two-level CTT (level 1 and level 2). When generating a new semantic graph, the new metapath is stored in the Candidate Register, and the CTT pointer (CP) begins searching from Level 1.
The data in the Candidate Register is then matched with the data at the CP. If the match is complete and the Next Pointer (Next P.) in the CP is empty, it indicates that the node is a leaf node. At this point, the Callback Pointer (Callback P.) is used to search, and the metapath from level 1 to the current node is used as an element for generating the semantic graph. If Next P. is not empty, the CP points to the position of Next P. to continue matching subsequent nodes until the Candidate Register matching is completed. The new semantic graph will also be stored in the CTT buffer.

\subsection{Graph Restructurer}
This section presents the details of the Graph Restructurer and its hardware implementation. We first identify an opportunity to alleviate the low cache hit rate and buffer thrashing issues. To capitalize on this, we propose the graph decoupling and recoupling algorithms, which are mapped to the decoupler and recoupler in the hardware design, respectively, to optimize the layout of the semantic graph and enhance the community structure of graph data.

\subsubsection{Opportunity for Graph Topology Optimization}\label{sec:opportunity}

As Section~\ref{sec:background} mentions, HGNNs do not directly operate on the original HetG. 
Instead, the original HetG is processed to build several directed and bipartite semantic graphs~\cite{OGB}.
In a bipartite graph, the vertex set can be divided into two non-empty parts such that all edges are between these two parts, with no edge within either part.

In graph theory, a bipartite graph has the characteristic of producing a maximal matching, which is the largest set of edges that connect unique vertices. The bipartite graph can establish a corresponding set of vertices from this set of edges, ensuring that every edge in the bipartite graph connects to at least one vertex within this defined set. We refer to this subset of vertices as the \textbf{graph backbone} in this paper, and it facilitates the classification of all vertices into four distinct parts:
\begin{itemize}
    \item $Src_{in}$: Source vertices included in the backbone. 
    \item $Src_{out}$: Source vertices excluded from the backbone. 
    \item $Dst_{in}$: Destination vertices included in the backbone. 
    \item $Dst_{out}$: Destination vertices excluded from the backbone.
\end{itemize}

By definition, there is no such edge whose endpoints are both outside the graph backbone, which derives the non-connectivity between $Src_{out}$ and $Dst_{out}$, and further a partition of the original graph into three distinct subgraphs.
These subgraphs demonstrate specific characteristics: all vertices connect to $Src_{out}$ belong to $Dst_{in}$, while all vertices connect to $Dst_{out}$ belong to $Src_{in}$. 

By eliminating irrelevant vertices from each subgraph and strategically scheduling the order of subgraph execution, more data can be reused in the buffer, effectively mitigating the problem of buffer thrashing.
Notably, these subgraphs exhibit a robust community structure, showcasing cohesive internal connections and relationships and thereby alleviating the issue of buffer thrashing.

\subsubsection{Graph Restructurer}\label{sec:graph_restructure_alg}
To harvest the opportunity for graph topology optimization, we propose a graph restructuring algorithm to reshape the semantic graphs, leveraging the inherent properties of bipartite graphs to improve the community structure.

\begin{figure}[!h] 
	\centering
	\includegraphics[width=0.48\textwidth]{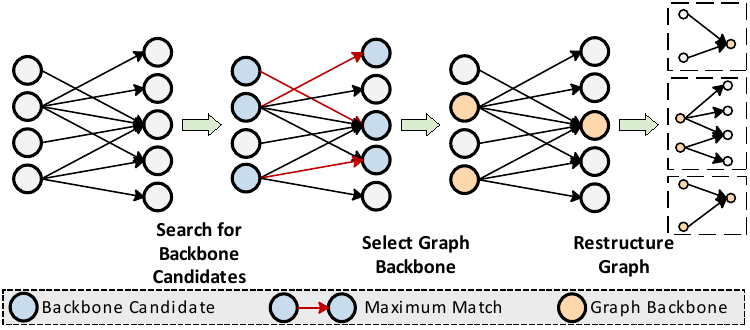}
	\caption{Toy example of graph restructuring algorithm.}
	\label{fig:toy_example}
\end{figure}

Fig.~\ref{fig:toy_example} offers a toy example to elucidate the workflow of the graph restructuring algorithm. 
This algorithm unfolds in two steps: graph decoupling and graph recoupling. In the former step, the maximum matching algorithm is employed to identify graph backbone candidates, effectively decoupling the semantic graph into a distinct edge group. In the latter step, the backbone is selected from this discrete edge group to reassemble the graph into three subgraphs, each characterized by a robust community structure. To implement these two steps, we propose a graph decoupling algorithm and a graph recoupling algorithm, respectively.

\textbf{Graph Decoupling Algorithm.}
The graph decoupling focuses on discovering the maximum matching within semantic graphs to identify graph backbone candidates, as illustrated in Algorithm~\ref{alg:decouple_algorithm}. Essentially, it draws inspiration from the Hungarian Algorithm~\cite{Hungarian_algorithm}. To optimize its execution, we design customized hardware by leveraging FIFO and hash table functionalities, as detailed in Section~\ref{sec:architecture}.

This algorithm begins by initializing all vertices without a match (line 2), subsequently scanning each vertex $u$ to identify its neighbor $v$ and commence the search.
Initially, for each neighbor $v$ of $u$, $u$ is added to the $Matchin\_FIFO$[$v$] (line 12) for temporary storage. If $v$ is unmatched, the algorithm records vertices $u$ and $v$ as a matched pair and frees the previous match associated with $u$ through iterative steps (lines 14-18).
In instances where all of $u$'s neighbors are matched, the vertices matched to $u$'s neighbors are pushed into $Search\_{List}$ (lines 22-26) to find another match, leaving the match for $u$.
After searching for the maximum match, the final matches are stored in $Match\_Pair$. Additionally, the matching vertices are recorded as graph backbone candidates, termed as $M$.

\begin{algorithm}[!htbp]
    \SetAlgoLined
    \footnotesize
    \caption{\textbf{Graph Decoupling Algorithm}}
    \label{alg:decouple_algorithm}
    \KwIn{$G^\mathcal{P}$: The input semantic graph;}
    \KwOut{$Match\_Pair$: Backbone candidate list;}
    $Match\_Pair$, $Search\_List$ = \{ \};\\
    Clear all $Matching\_FIFO$;\\
    \For{each vertex $n$ in $G^\mathcal{P}$}{
        \If{$Match\_Pair$[$n$] \textless 0}{
            Push $n$ to $Search\_List$;\\
            \While{$Search\_List$ is not empty}{
                Pop $u$ from $Search\_List$;\\
                \For{each neighbor $v$ of $u$}{
                    \If{$v$ is visited}{
                       continue;\\
                    }
                    Push $u$ to $Matching\_FIFO$[$v$];\\
                    \If{$Match\_Pair$[$v$] \textless 0}{
                        \While{$Match\_Pair$[$u$] \textgreater 0}{
                            $Matching\_FIFO$[$Match\_Pair$[$u$]].pop();\\
                            Change $Match\_Pair$;\\
                            $u$ = $Match\_Pair$[$Match\_Pair$[$u$]];\\
                        }
                        break while;\\
                    }
                }
                \If{$Match\_Pair$[$v$] \textless 0}{
                    \For{each matched neighbor $u$ of $v$}{
                        Push $Match\_Pair$[$u$] to $Search\_List$;\\
                    }
                }
            }
        }
    }
    return $Match\_Pair$;\\
\end{algorithm}

\textbf{Graph Recouping Algorithm.}
The graph recoupling aims to select the graph backbone from the candidates and generate subgraphs, as shown in Algorithm~\ref{alg:recouple_algorithm}.

Initially, the algorithm initiates backbone selection. It commences by exploring all matched vertices categorized by source and destination. For each matched source vertex, the algorithm identifies its unmatched neighbors and puts them to $Dst_{out}$ while putting itself to $Src_{in}$. If all its neighbors are matched, the algorithm will just skip to the next matched source vertex. Also, all matched destination vertices will be examined after the matched source vertices for the same procedure. After all, all left vertices are then classified to $Src_{out}$ or $Dst_{out}$, according to whether they belong to $V_{src}$ or $V_{dst}$.
Once the backbone is selected, subgraphs are generated for the subsequent execution.

\begin{algorithm}[!h]
    \SetAlgoLined
    \footnotesize
    \caption{\textbf{Graph Recoupling Algorithm}}
    \label{alg:recouple_algorithm}
    \KwIn{$G^\mathcal{P}$: The input graph;  $M$: Backbone candidate vertex set;}
    \KwOut{$G_{s_1}^{\mathcal{P}}$, $G_{s_2}^{\mathcal{P}}$, $G_{s_3}^{\mathcal{P}}$: The subgraphs generated from original semantic graph.}
    $Src_{in}, Src_{out}, Dst_{in}, Dst_{out}$ = \{ \};\\
    $S\leftarrow V_{src}\cap M, \overline{S}\leftarrow V_{src}\setminus M, T\leftarrow V_{dst}\cap M, \overline{T}\leftarrow V_{dst}\setminus M$; \\
    \For{each $v$ in $S$}{
        $X_v \leftarrow N(v)\cap \overline{T}$; \\
        \If{$X_v$ is not $\varnothing$}{
            Push $v$ to $Src_{in}$;\\
            Push $X_v$ to $Dst_{out}$;\\
        }
    }
    \For{each $u$ in $T$}{
        $X_u \leftarrow N(u)\cap \overline{S}$; \\
        \If{$X_u$ is not $\varnothing$}{
            Push $u$ to $Dst_{in}$;\\
            Push $X_u$ to $Src_{out}$;\\
        }
    }
    Push the other source vertices to $Src_{out}$;\\
    Push the other destination vertices to $Dst_{out}$;\\
    $G_{s_1}^{\mathcal{P}}$, $G_{s_2}^{\mathcal{P}}$, $G_{s_3}^{\mathcal{P}}$ = \textbf{GenerateGraph}($Src_{in}$, $Src_{out}$, $Dst_{in}$, $Dst_{out}$);\\
    return $G_{s_1}^{\mathcal{P}}$, $G_{s_2}^{\mathcal{P}}$, $G_{s_3}^{\mathcal{P}}$;
\end{algorithm}

\subsubsection{Hardware Implementation}\label{sec:architecture}
In this section, we detail the algorithm mapping to the hardware and the design of the Graph Restructurer.

\begin{figure}[!t] 
	\centering
	\includegraphics[width=0.3\textwidth]{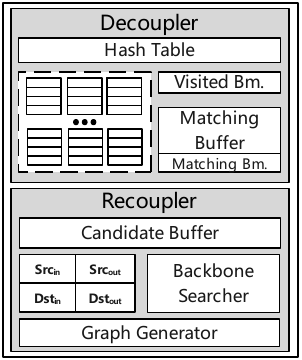}
	\caption{The overview of the Graph Restructurer architecture.}
	\label{fig:restructuring_module}
\end{figure}

Fig.~\ref{fig:restructuring_module} offers the overview of the Graph Restructurer, which primarily comprises two modules: Decoupler and Recoupler. 
The Decoupler undertakes graph decoupling and is constructed with components of the Hash Table, FIFOs, bitmaps (Bm.), and buffers. 
On the other hand, the Recoupler is tasked with executing graph recoupling and consists of the Backbone Searcher, a collection of FIFOs, and the Graph Generator.

\begin{figure}[!h] 
	\centering
	\includegraphics[width=0.3\textwidth]{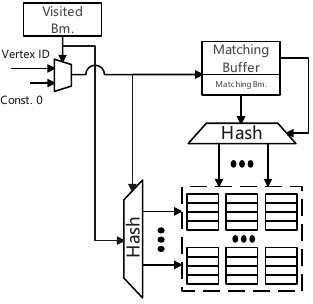}
	\caption{Micro-architecture of Decoupler.}
	\label{fig:decoupler_micro_architecture}
\end{figure}

\textbf{Workflow of Decoupler.} In each execution epoch, the topology of the original semantic graph is received and passed on to the hash table for FIFO allocation.
The FIFOs, organized in a set-associative manner, store matched pairs and waiting lists allocated to specific vertices. 
As illustrated in Fig.~\ref{fig:decoupler_micro_architecture}, during each cycle, source vertices are dispatched to their respective FIFOs, automatically triggering a pop operation for FIFOs if the match condition changes. 
The Matching Buffer stores replaced FIFO data. Upon identifying all matching edges, the resulting backbone candidates are stored in each FIFO and sent to the Candidate Buffer.

\begin{figure}[!h] 
	\centering
	\includegraphics[width=0.34\textwidth]{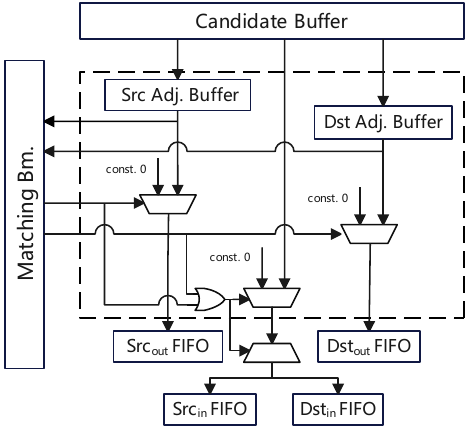}
	\caption{Micro-architecture of Recoupler.}
	\label{fig:recoupler_micro_architecture}
\end{figure}

\textbf{Workflow of Recoupler.} The identified candidates are then forwarded to the Backbone Searcher, as highlighted in Fig.~\ref{fig:recoupler_micro_architecture}. 
During this step, the Candidate Buffer transmits the backbone candidates to the Backbone Searcher to identify the graph backbone, following Algorithm~\ref{alg:recouple_algorithm}.
Initially, each candidate is directed to the adjacency list buffers including the Src Adj. Buffer and Dst Adj. Buffer to obtain their respective neighbors. Subsequently, all obtained neighbors are checked in the Matching Bm. If any neighbors are not found in the Matching Bm., the candidate is sent to either $Src_{in}$ or $Dst_{in}$ FIFOs, depending on its origin, and the corresponding neighbors are dispatched to the $Src_{out}$ or $Dst_{out}$ FIFOs.
The Graph Generator creates the subgraphs from these four designated buffers and forwards them for subsequent HGNN execution.

\section{Experimental Setup}\label{sec:exp_setup}
In this section, we introduce the experimental setup used to assess the performance of SiHGNN.

\subsection{Evaluation Methodology}

To better demonstrate the performance of SiHGNN during HGNN execution, we select the state-of-the-art HGNN accelerator HiHGNN~\cite{HiHGNN} to be used in conjunction with SiHGNN. HiHGNN utilizes a hybrid architecture that includes a systolic array module for matrix-vector multiplication and a SIMD module for performing element-wise operations during HGNN execution, covering the FP, NA, and SF sub-stages. It is important to note that SiHGNN can also be integrated with other future HGNN accelerators for enhanced performance.
In the following section, SiHGNN will refer to the combination of SiHGNN and HiHGNN.
The performance metrics of SiHGNN are evaluated using the following tools.

\textbf{Architecture Simulator.}
We implement SiHGNN in a cycle-level accurate simulator to measure execution time in the number of cycles. We also design a detailed cycle-accurate on-chip memory model and integrate the Ramulator~\cite{ramulator} for FIFOs, buffers, and memory simulation. This simulator models the micro-architectural behaviors of each module and the hardware datapath.

\textbf{CAD Tools.}
We implement an RTL version of each hardware module and synthesize it to evaluate the area, energy consumption, and latency. We use the Synopsys Design Compiler with the TSMC 12 \textit{nm} standard VT library for the synthesis and estimate the power consumption using Synopsys PrimeTime PX.

\textbf{Memory Measurements.}
We estimate the buffer area, energy consumption, and access latency using Cacti~\cite{CACTI}. 
Since Cacti only supports 32 nm technology, we use four different scaling factors to convert them to 12 nm technology.
The access latency and energy of HBM 1.0 are simulated by the Ramulator~\cite{ramulator} and estimated with 7 pJ/bit~\cite{7pj} as HiHGNN, respectively.

\subsection{Models and Datasets}

We conduct the experiments on three different models including RGCN~\cite{RGCN}, RGAT~\cite{RGAT}, and Simple-HGN~\cite{Simple-HGN}. 
We leverage three datasets, ACM, DBLP, and IMDB, which are commonly used in the HGNN research community~\cite{DGL, understand_HGNN, HiHGNN}. The details of the datasets are listed in Table~\ref{tb:datasets}.
The raw features of each vertex are provided by the respective datasets. 
In the following experiments, we generate semantic graphs with relations ranging from two to nine hops.

\begin{table}[!t]
\centering
\footnotesize
\caption{Information of HetG Datasets.} \label{tb:datasets}
\renewcommand\arraystretch{0.6}
\setlength\tabcolsep{2pt}%
\resizebox{0.45\textwidth}{!}{
\begin{tabular}{cccc}
    \toprule
    \textbf{Dataset} &
      \makebox[0.1\textwidth][c]{\textbf{\#Vertex}} &
      \makebox[0.1\textwidth][c]{\textbf{\#Feature}} &
      \makebox[0.1\textwidth][c]{\textbf{Relations}} \\ \midrule
    \multirow{4}{*}{IMDB}   & movie (M): 4932       & M: 3489   & \multirow{4}{*}{\begin{tabular}[c]{@{}c@{}}A $\rightarrow$ M M $\rightarrow$ A\\ K $\rightarrow$ M M $\rightarrow$ K\\ D $\rightarrow$ M M $\rightarrow$ D\end{tabular}}\\
                            & director (D): 2393    & D: 3341   &          \\
                            & actor (A): 6124       & A: 3341   &          \\
                            & keyword (K): 7971     & K: ---    &          \\ \midrule
    \multirow{4}{*}{ACM}    & paper (P): 3025       & P: 1902   & \multirow{4}{*}{\begin{tabular}[c]{@{}c@{}}T $\rightarrow$ P P $\rightarrow$ T\\ S $\rightarrow$ P P $\rightarrow$ S\\ P $\rightarrow$ P -P $\rightarrow$ P \\ A $\rightarrow$ P P $\rightarrow$ A\end{tabular}}\\
                            & author (A): 5959      & A: 1902   &          \\
                            & subject (S): 56     & S: 1902   &          \\
                            & term (T): 1902        & T: ---    &          \\ \midrule
    \multirow{4}{*}{DBLP}   & author (A): 4057      & A: 334    & \multirow{4}{*}{\begin{tabular}[c]{@{}c@{}}A $\rightarrow$ P P $\rightarrow$ A\\ V $\rightarrow$ P P $\rightarrow$ V\\ T $\rightarrow$ P P $\rightarrow$ T\end{tabular}}\\
                            & paper (P): 14328       & P: 4231   &          \\
                            & term (T): 7723       & T: 50     &          \\
                            & venue (V): 20       & V: ---    &          \\ \bottomrule
\end{tabular}
}
\scriptsize
\end{table}

\subsection{Baseline Platforms}
To compare the performance of SiHGNN with the state-of-the-art work, we implement all HGNN models using the state-of-the-art framework DGL 1.0.2 \cite{DGL} and evaluate them on both an NVIDIA T4 GPU and an NVIDIA A100 GPU, utilizing NVIDIA Nsight Compute for analysis. All models are configured with the same number of hidden units \{64\} and layers \{3, 3, 2\} for \{R-GAT, R-GCN, S-HGN\}, respectively.
Additionally, the same HGNN models are implemented in SiHGNN, maintaining identical configurations for hidden units and layers as used on the GPUs. Table~\ref{tb:baselines} lists the configurations for HiHGNN and SiHGNN solely.

\begin{table}[!t]
\centering
\caption{Platform Details of HiHGNN and SiHGNN.}\label{tb:baselines}
\renewcommand\arraystretch{1.2}
\setlength\tabcolsep{2pt}%
\resizebox{0.47\textwidth}{!}{
\begin{tabular}{ccc}
\toprule
 & \textbf{HiHGNN} & \textbf{SiHGNN (Sole)} \\ \midrule
\begin{tabular}[c]{@{}c@{}}Peak Performance\end{tabular} & 

\multicolumn{1}{c}{\begin{tabular}[c]{@{}c@{}}16.38 TFLOPS, 1.0 GHz\end{tabular}} & --- \\ \midrule
On-chip Buffer & \begin{tabular}[c]{@{}c@{}}2.44 MB (FP-Buf),\\ 14.52 MB (NA-Buf),\\ 0.12 MB (SA-Buf),\\ 0.38 MB (Att-Buf)\end{tabular} & \begin{tabular}[c]{@{}c@{}} 5 KB (CTT Buf),\\ 168 KB (Decoupler),\\ 480 KB (Recoupler)\end{tabular} \\ \midrule
Off-chip Memory &
\multicolumn{1}{c}{\begin{tabular}[c]{@{}c@{}}512 GB/s, HBM 1.0\end{tabular}} & --- \\ \bottomrule
\end{tabular}
}
\end{table}

\section{Experiment Results}
In this section, we compare SiHGNN with state-of-the-art execution platforms and provide a detailed optimization analysis. We demonstrate the advantages of SiHGNN in terms of speedup, memory access, and other key metrics by comparing it against a state-of-the-art software framework operating on both the NVIDIA T4 GPU and A100 GPU, as well as the state-of-the-art HGNN accelerator HiHGNN.

\subsection{Overall Results}

\subsubsection{Speedup}
\begin{figure}[!h] 
	
	\centering
	\includegraphics[width=0.48\textwidth]{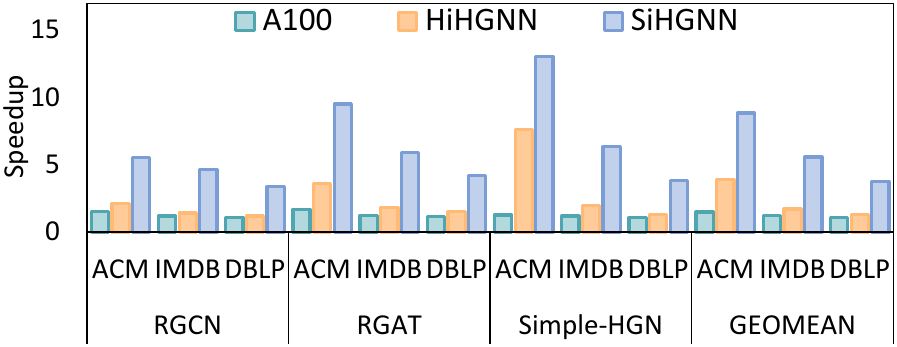}
	\caption{Overall speedup to T4 GPU.}
	\label{fig:speedup}
	
\end{figure}

Fig.~\ref{fig:speedup} shows the overall speedup of the A100 GPU, HiHGNN, and SiHGNN compared to the T4 GPU. The last set of bars, labeled GEOMEAN, represents the geometric mean across all HGNN models. We generate four different semantic graphs for each dataset using three- and four-hop metapaths, which are the most commonly used in real-world scenarios. 
SiHGNN achieves an average speedup of 6.15$\times$, 4.86$\times$, and 2.95$\times$ compared to T4 GPU, A100 GPU, and HiHGNN, respectively.

The performance improvement stems from two primary aspects. Firstly, SiHGNN introduces the Semantic Graph Builder to accelerate the SGB stage. This builder utilizes the CTT to guide the generation of semantic graphs, eliminating redundant computations and memory accesses, thereby enhancing the overall performance. 
Secondly, SiHGNN incorporates a graph restructuring algorithm, mapped to the Graph Restructurer. This module transforms the structure of semantic graphs into a more community-focused format, which reduces buffer replacements and improves performance. 
Additionally, the pipeline design within SiHGNN ensures uninterrupted utilization of intermediate results, further minimizing buffer replacements and contributing to the performance boost.

\subsubsection{Area and Power Overhead}
\begin{figure}[!h] 
	\centering
	\includegraphics[width=0.48\textwidth]{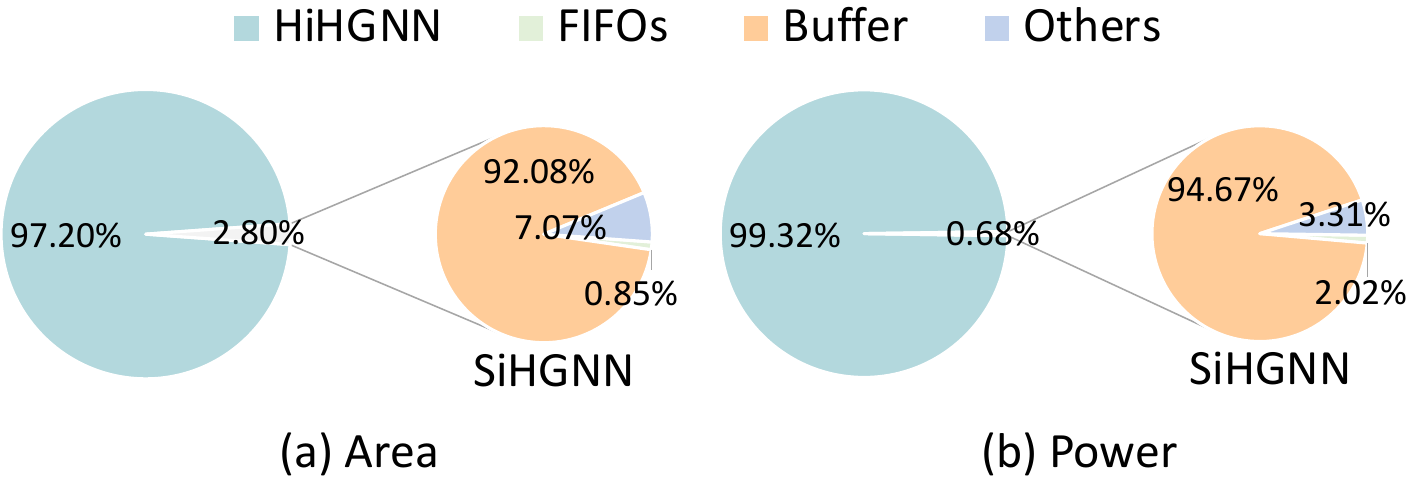}
	\caption{Area and power of HiHGNN and SiHGNN.}
	\label{fig:power_area_overhead}
\end{figure}

Compared to the A100 GPU at 7 $nm$, the integrated SiHGNN under TSMC 12 $nm$ consumes only 5\% of the power (i.e., 12.09 $W$ vs. 250 $W$) and occupies just 2.6\% of the area (i.e., 21.7 $mm^2$ vs. 826 $mm^2$).
Fig.~\ref{fig:power_area_overhead} further shows the area and power breakdown for SiHGNN when integrated with HiHGNN. The results reveal that SiHGNN contributes minimally to overall resource consumption, accounting for only 2.80\% (i.e., 0.61 $mm^2$) of the total area and 0.68\% (i.e., 82.2 $mW$) of the total power when integrated with HiHGNN under TSMC 12 $nm$ technology. This confirms that the overhead introduced by the Semantic Graph Builder and the Graph Restructurer within SiHGNN is negligible, ensuring that it does not impose a significant burden on the overall system performance or resource utilization.
Consequently, this makes SiHGNN a highly efficient and cost-effective addition as an accelerator frontend, offering enhanced performance with minimal resource trade-offs.

\subsection{Effects of Proposed Optimizations}
Detailed evaluations are conducted to provide deeper insights into the proposed optimizations, focusing on the benefits of the Semantic Graph Builder in the SGB stage and the improvements brought by the Graph Restructurer in the GFP stage. 
We compare the effectiveness of two configurations: (1) with and without the Semantic Graph Builder during the SGB stage, and (2) with and without the Graph Restructurer during the GFP stage.

\subsubsection{Effects of the Semantic Graph Builder}

To demonstrate the impact of the Semantic Graph Builder, we evaluate the speedup of semantic graph generation and provide a detailed analysis of computation and memory access reductions during the SGB stage. Semantic graphs ranging from three to nine hops are generated using the ACM, DBLP, and IMDB datasets. The results are compared on SiHGNN, both with and without the integration of the Semantic Graph Builder.

\begin{figure}[!t] 
	\centering
	\includegraphics[width=0.48\textwidth]{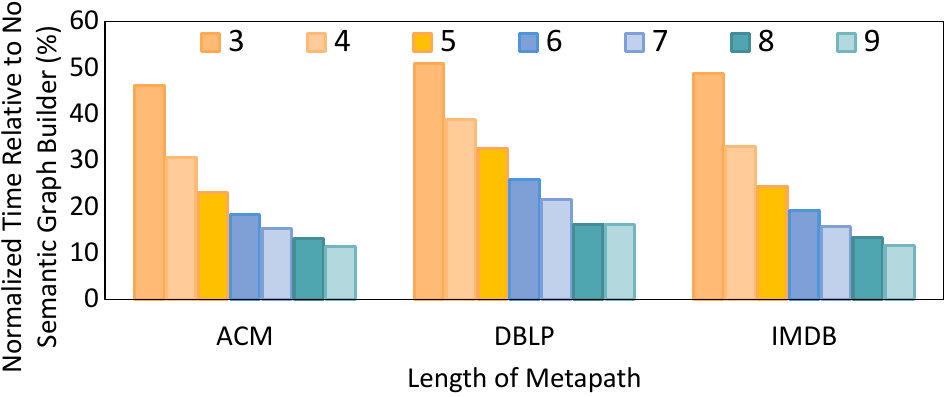}
	\caption{A comparison of the speedup achieved during semantic graph building in the SGB stage with and without the Semantic Graph Builder.}
	\label{fig:speedup_SGB}
\end{figure}

\textbf{Speedup.} Fig.~\ref{fig:speedup_SGB} compares the speedup in the SGB stage on SiHGNN between with and without the Semantic Graph Builder, focusing on semantic graph lengths ranging from three to nine. Guided by the CTT, semantic graph generation in SiHGNN with the Semantic Graph Builder achieves significant performance improvement, particularly with longer metapaths.
These gains are driven by the reduction of redundant computations and memory access. As the length of semantic graphs increases, the Semantic Graph Builder optimizes the generation of candidate semantic graphs by leveraging shorter existing semantic graphs, leading to substantial reductions in computation and memory access, and thereby enhancing performance.

\begin{figure}[!h] 
	\centering
	\includegraphics[width=0.48\textwidth]{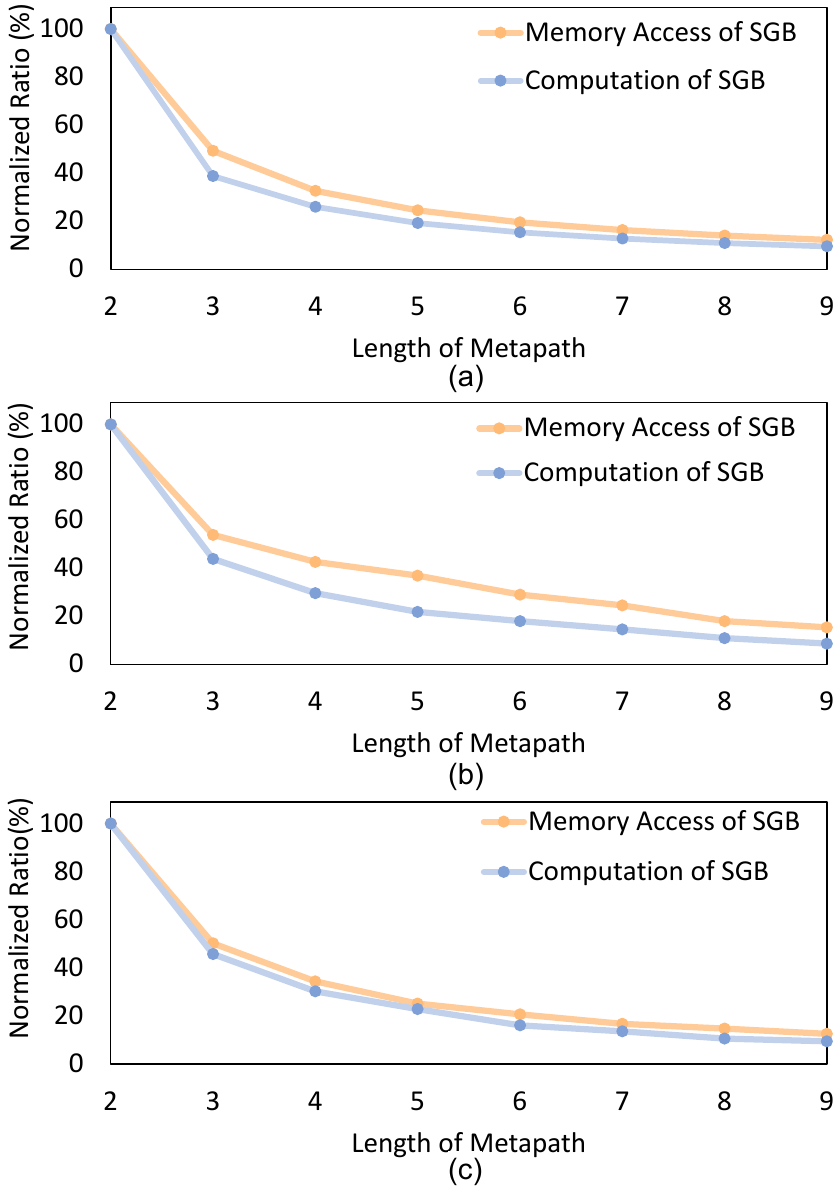}
	\caption{The computation reduction and the memory accesses reduction in the SGB stage with and without the Semantic Graph Builder. (a) The results on the ACM dataset; (b) The results on the DBLP dataset; (c) The results on the IMDB dataset.}
	\label{fig:reduction_SGB}
\end{figure}

\textbf{Redundancy Reduction.} We conduct further experiments to quantify the reductions in computation and memory access during the SGB stage. Using the same experimental setup, we generate semantic graphs ranging from three to nine hops across three datasets.
As shown in Fig.~\ref{fig:reduction_SGB}, the Semantic Graph Builder achieves significantly greater reductions in both computation and memory access, with the improvement becoming more pronounced as the metapath length increases. 
This enhancement is primarily driven by the CTT, which optimizes the generation for each semantic graph by leveraging the existing semantic graphs, leading to substantial reductions in computation and memory access during the SGB stage.

\subsubsection{Effects of the Graph Restructurer}
In this section, we illustrate the improvements in the GFP stage provided by the Graph Restructurer. 
To isolate the effects of semantic graph generation, we conduct experiments using one-hop relations to evaluate speedup, DRAM access reduction, and bandwidth utilization. We compare SiHGNN against all baselines, including the GPU A100, GPU T4, and HiHGNN.

\begin{figure}[!h] 
	\centering
	\includegraphics[width=0.48\textwidth]{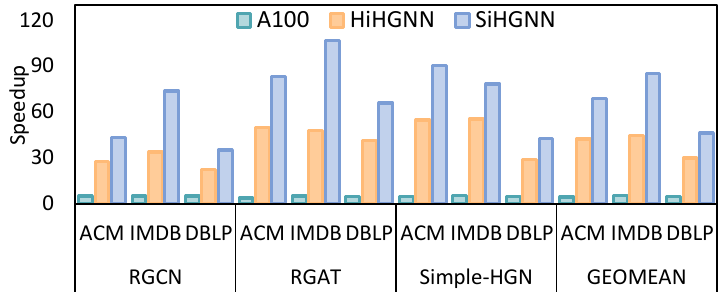}
	\caption{Speedup in the GFP stage to GPU T4.}
	\label{fig:speedup_inference}
\end{figure}

\textbf{Speedup.} Fig.~\ref{fig:speedup_inference} shows the speedup of SiHGNN with other baselines in the GFP stage. In general, SiHGNN achieves an average speedup of 68.8$\times$,
14.6$\times$ and 1.78$\times$ compared to T4 GPU, A100 GPU, and HiHGNN in the GFP stage, respectively. 
The Graph Restructurer facilitates the transformation of graph structures, evolving from a pattern that generates numerous random accesses to an organized form that exhibits community locality. This restructuring reduces buffer replacements and increases the buffer hit rate, thereby enhancing overall performance. 
Additionally, the pipeline between the Decoupler, Recoupler, and accelerator ensures uninterrupted utilization of intermediate results, further minimizing buffer replacements and leading to a significant performance boost.

\begin{figure}[!h] 
	
	\centering
	\includegraphics[width=0.48\textwidth]{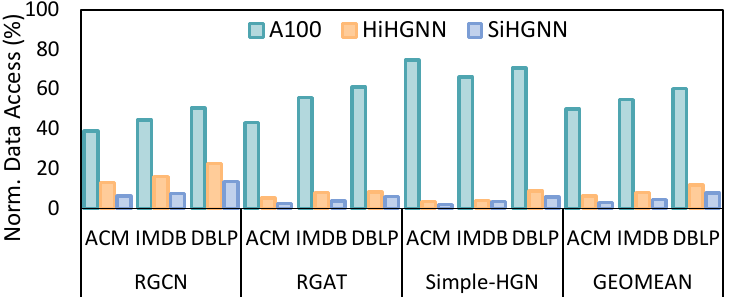}
	
	\caption{Normalized DRAM access to T4 GPU.}
	\label{fig:DRAM_access}
	
\end{figure}

\textbf{DRAM Accesses.} To analyze the source of performance improvement in the GFP stage, Fig.~\ref{fig:DRAM_access} presents the normalized DRAM access for SiHGNN and other baselines during runtime, relative to the T4 GPU. 
SiHGNN reduces the need for DRAM access by enhancing the community structure of semantic graphs and improving the buffer hit rate of frequently accessed data. 
Specifically, during the GFP stage, SiHGNN reduces data access to just 4.8\%, 8.7\%, and 57.1\% compared to the T4 GPU, A100 GPU, and HiHGNN, respectively.

From the model perspective, the RGAT and Simple-HGN models, which adopt an attention-based NA sub-stage, greatly benefit from the reuse of intermediate results. In contrast, the FP sub-stage dominates the execution in the R-GCN model, making significant DRAM access to raw features inevitable.
This result confirms that, with the assistance of SiHGNN, HiHGNN significantly reduces the number of DRAM accesses, validating the source of the observed speedup.

\begin{figure}[!h] 
	\centering
	\includegraphics[width=0.48\textwidth]{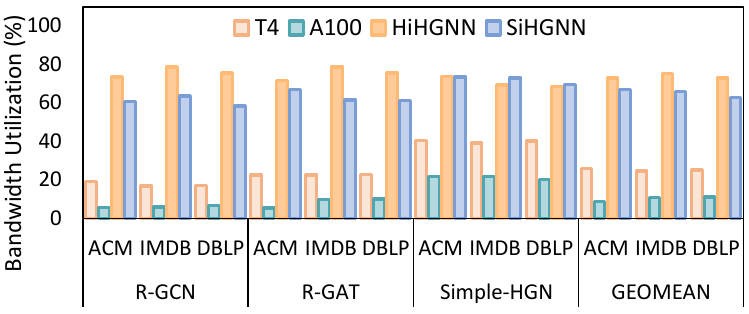}
	\caption{DRAM bandwidth utilization.}
	\label{fig:bandwidth_utilization}
\end{figure}

\textbf{DRAM Bandwidth.} Fig.~\ref{fig:bandwidth_utilization} shows the utilization of the DRAM bandwidth of SiHGNN and the baselines in the GFP stage. 
SiHGNN demonstrates 2.58$\times$ and 6.35$\times$ improvement on average in the utilization of DRAM bandwidth compared with T4 GPU and A100 GPU, respectively.
The limited bandwidth utilization in GPUs is due to random accesses to projected features, attention coefficients, edge embeddings, and other intermediate results. In contrast, SiHGNN minimizes random DRAM accesses by carefully managing and storing these intermediate results in on-chip buffers. Compared to HiHGNN, SiHGNN significantly reduces DRAM accesses through graph restructuring, though this improvement slightly impacts overall bandwidth utilization due to increased demands on compute resources.

\section{Related Work}
Given the remarkable learning capacity of GNNs with graph data, GNN accelerators have garnered significant attention from the architecture community~\cite{HyGCN,igcn,FlowGNN,rrgat, AWB_GCN,GCNAX,multigcn_inter_node_communication}. HyGCN~\cite{HyGCN} introduces a hybrid architecture with dedicated modules for aggregation and combination, leveraging inter-stage fusion to enhance overall performance. AWB-GCN~\cite{AWB_GCN} addresses the workload imbalance issues caused by the power-law distribution of non-zeros in the adjacency matrix by proposing a workload autotuning technique. Both HyGCN and AWB-GCN focus on Graph Convolutional Networks (GCNs) and their variants, optimizing hardware datapath and scheduling for these specific models.
MultiGCN~\cite{multigcn_inter_node_communication} introduces an efficient multi-node GCN accelerator for large-scale graphs, strategically designed to minimize redundant network transmissions and off-chip memory accesses by leveraging a trade-off between network latency and bandwidth. This is accomplished through a topology-aware multicast mechanism combined with a scatter-based round execution approach, effectively addressing the challenges posed by irregular communication patterns and network bandwidth constraints in multi-node GCN acceleration.
I-GCN~\cite{igcn} presents another approach to GCN acceleration, introducing the ``islandization'' method to improve data locality in GCNs. This technique identifies clusters of vertices with strong internal connections but weak external connections. However, this method is not suitable for semantic graphs, which are directed bipartite graphs. The properties of such graphs cause this method to degrade into a process focused solely on finding the vertex with the largest degree, making it less effective for handling the complexities of semantic graphs.

Previous efforts~\cite{HiHGNN, MetaNMP, ADE-HGNN} have proposed several accelerators for HGNN acceleration. HiHGNN~\cite{HiHGNN} strategically schedules the execution order of semantic graphs based on their similarity to exploit data reusability across different semantic graphs. However, it primarily focuses on dynamic data optimization and overlooks the SGB stage, which is crucial for inductive graph learning.
MetaNMP~\cite{MetaNMP} pioneers DIMM-based near-memory processing for HGNNs, employing a cartesian-like product paradigm to dynamically generate metapath instances and aggregate vertex features from the starting vertex along these metapaths. However, MetaNMP mainly targets HGNN models that use intra-metapath aggregation~\cite{MAGNN} to aggregate vertex features during the feature projection stage. 
ADE-HGNN~\cite{ADE-HGNN} introduces a runtime pruning approach based on a min-heap to discard insignificant neighbors and presents a novel execution flow based on operation fusion to effectively amortize pruning overhead. However, ADE-HGNN primarily accelerates the attention mechanism in HGNNs.

\section{Conclusion}
In this work, we highlight the often-overlooked potential for improving performance in HGNNs by leveraging properties of semantic graphs during both the SGB stage and the GFP stage. To address this, we introduce SiHGNN, a lightweight hardware accelerator frontend specifically designed to enhance the execution of HGNNs.
SiHGNN incorporates a tree-based Semantic Graph Builder for efficient semantic graph generation and a Graph Restructurer to optimize the layout of semantic graphs. Our experimental results demonstrate that SiHGNN achieves significant improvements in performance and energy efficiency compared to GPUs and the state-of-the-art HGNN accelerator, HiHGNN.

\ifCLASSOPTIONcompsoc
  \section*{Acknowledgments}
\else
  \section*{Acknowledgment}
\fi

This work was supported by National Key Research and Development Program (Grant No. 2022YFB4501400), the National Natural Science Foundation of China (Grant No. 62202451), CAS Project for Young Scientists in Basic Research (Grant No. YSBR-029), and CAS Project for Youth Innovation Promotion Association.

\ifCLASSOPTIONcaptionsoff
  \newpage
\fi



\bibliographystyle{IEEEtranS}
\bibliography{ref}

%


\vspace{-30pt}
\begin{IEEEbiography}[{\includegraphics[width=1in,height=1.25in,clip,keepaspectratio]{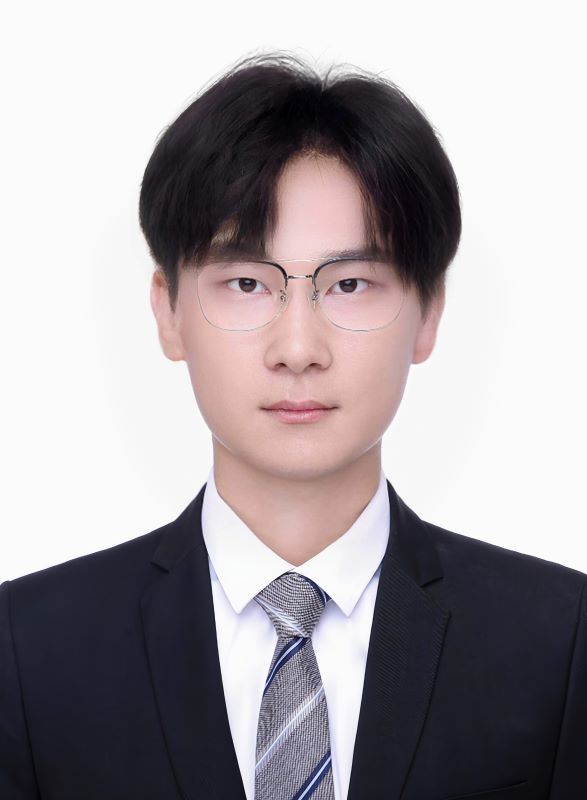}}]{Runzhen Xue}
received his B.E. degree from Shandong University, Qingdao, China in 2021. He is currently a Ph.D. candidate at the Institute of Computing Technology, Chinese Academy of Sciences, Beijing, China. His research interests include the hardware accelerator, high-performance computer architecture, and processor design space exploration.
\end{IEEEbiography}

\vspace{-30pt}
\begin{IEEEbiography}[{\includegraphics[width=1in, height=1.25in, clip, keepaspectratio]{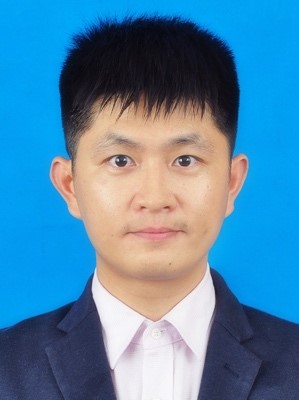}}] {Mingyu Yan} 
received his Ph.D. degree from the University of Chinese Academy of Sciences, Beijing, China in 2020. He is currently an associate professor at the Institute of Computing Technology, Chinese Academy of Sciences, Beijing, China. His current research interests include the graph processing algorithm, graph-based hardware accelerator, and high-throughput computer architecture.
\end{IEEEbiography}

\vspace{-30pt}
\begin{IEEEbiography}[{\includegraphics[width=1in,height=1.25in,clip,keepaspectratio]{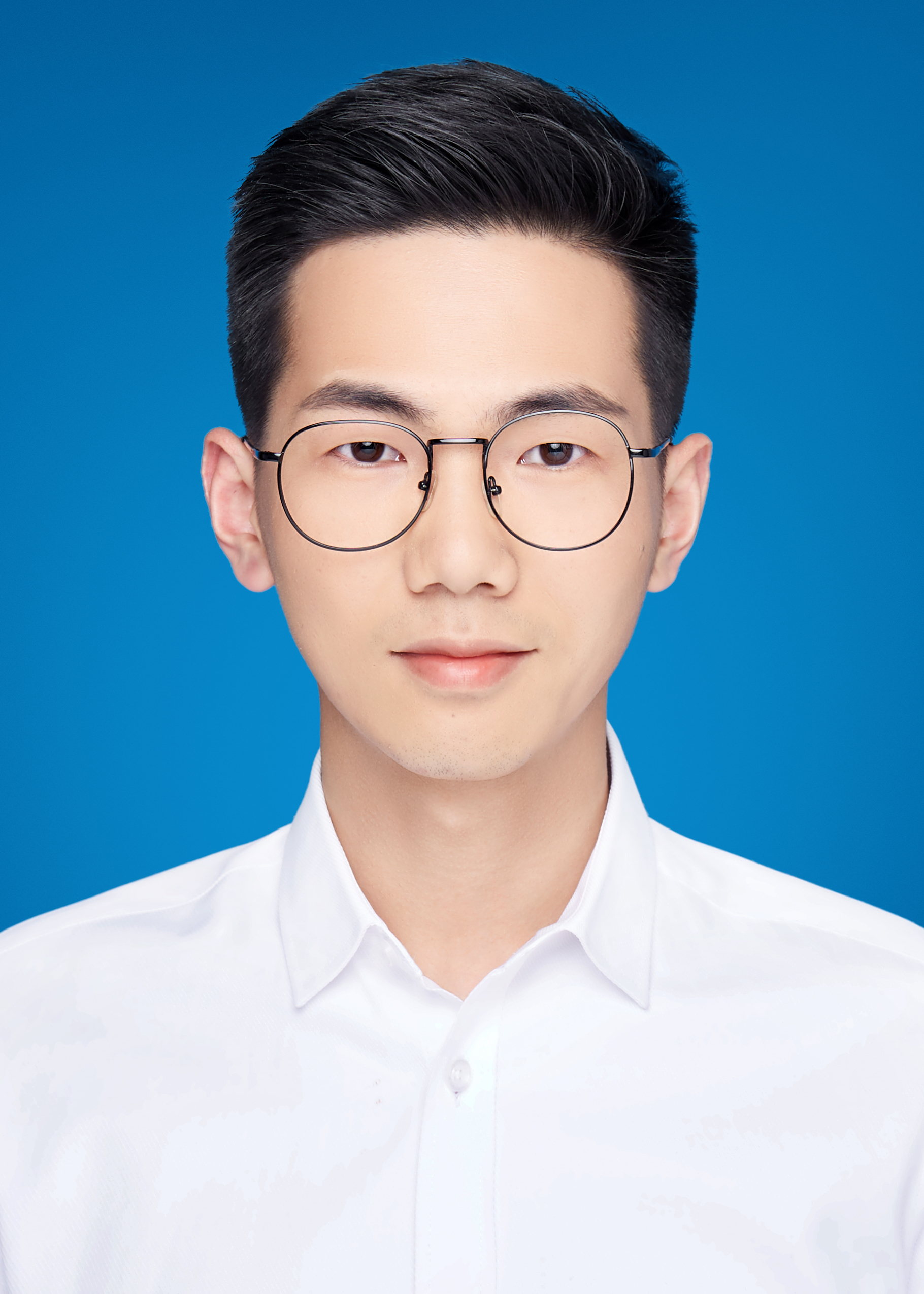}}]{Dengke Han}
He is currently a Ph.D. candidate at the Institute of Computing Technology, Chinese Academy of Sciences, Beijing, China. His current research interests include the graph-based hardware accelerator and high-throughput computer architecture.
\end{IEEEbiography}

\vspace{-30pt}
\begin{IEEEbiography}[{\includegraphics[width=1in,height=1.25in,clip,keepaspectratio]{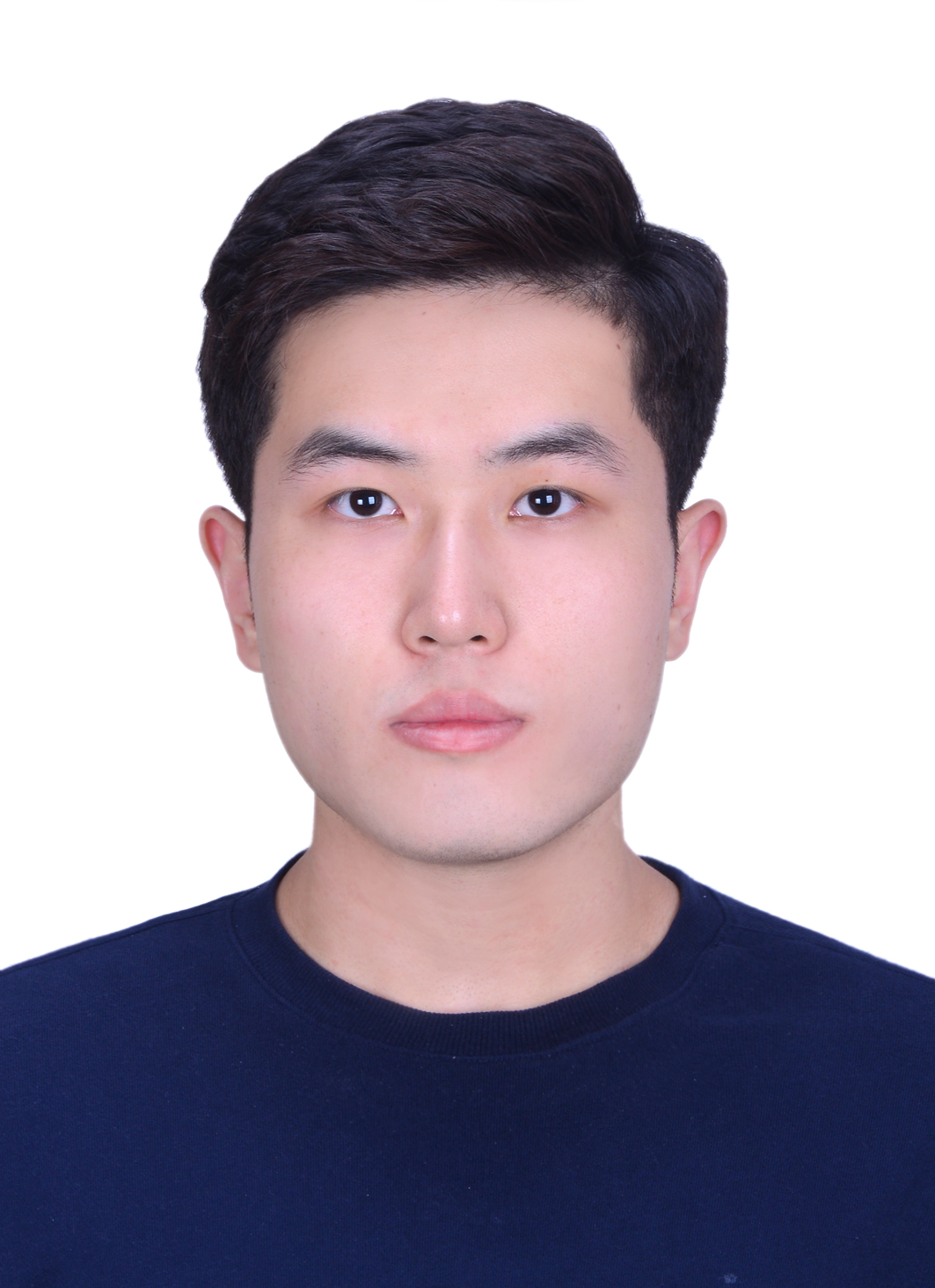}}]{Ziheng xiao}
received his M.S. degree from the Institute of Computing Technology, Chinese Academy of Sciences, Beijing, China, in 2024. His current research interests include graph-based hardware accelerators, domain-specific hardware architectures for cryptographic algorithms, and high-throughput computer architecture.
\end{IEEEbiography}

\vspace{-30pt}
\begin{IEEEbiography}[{\includegraphics[width=1in, height=1.25in, clip, keepaspectratio]{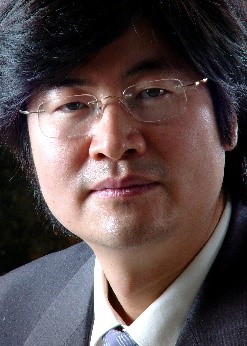}}] {Zhimin Tang} 
received his Ph.D. degree in computer architecture from the Institute of Computing Technology, Chinese Academy of Sciences, Beijing, in 1990. He is currently a professor and Ph.D. supervisor at the Institute of Computing Technology, Chinese Academy of Sciences, Beijing. His main research interests include high-throughput computer architecture, high-performance computer architecture, and multi \& many-core processor design. 
\end{IEEEbiography}

\vspace{-35pt}
\begin{IEEEbiography}[{\includegraphics[width=1in, height=1.25in, clip, keepaspectratio]{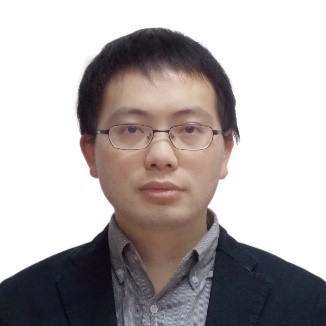}}] {Xiaochun Ye} 
received his Ph.D. degree in computer architecture from the Institute of Computing Technology, Chinese Academy of Sciences, Beijing, in 2010. He is currently a professor and Ph.D. supervisor at the Institute of Computing Technology, Chinese Academy of Sciences, Beijing. His main research interests include high-performance computer architecture and software simulation.
\end{IEEEbiography}

\vspace{-40pt}
\begin{IEEEbiography}[{\includegraphics[width=1in, height=1.25in, clip, keepaspectratio]{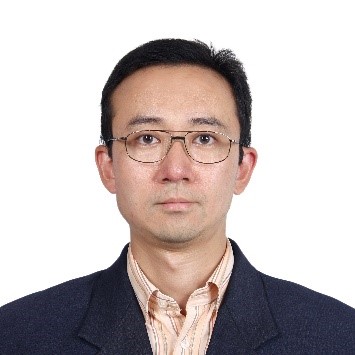}}] {Dongrui Fan} 
received his Ph.D. degree in computer architecture from the Institute of Computing Technology, Chinese Academy of Sciences, Beijing, in 2005. He is currently a professor and Ph.D. supervisor at the Institute of Computing Technology, Chinese Academy of Sciences, Beijing. His main research interests include high-throughput computer architecture and high-performance computer architecture. 

\end{IEEEbiography}

%





\end{document}